\documentclass[aps,prl,reprint,superscriptaddress]{revtex4-2}

\usepackage{graphicx}
\usepackage{dcolumn}
\usepackage{hyperref}
\usepackage{amsmath}

\begin{document}

\title{Synchronization of non-solitonic Kerr combs}

\author{Bok~Young~Kim}
\author{Jae~K.~Jang}
\author{Yoshitomo~Okawachi}
\affiliation{Department of Applied Physics and Applied Mathematics, Columbia University, New York, NY 10027, USA}
\author{Xingchen~Ji}
\affiliation{Department of Electrical Engineering, Columbia University, New York, NY 10027, USA}
\author{Michal~Lipson}
\author{Alexander~L.~Gaeta}
\email{a.gaeta@columbia.edu}
\affiliation{Department of Applied Physics and Applied Mathematics, Columbia University, New York, NY 10027, USA}
\affiliation{Department of Electrical Engineering, Columbia University, New York, NY 10027, USA}

\date{\today}

\begin{abstract}
  Synchronization is a ubiquitous phenomenon in nature that manifests as the spectral or temporal locking of coupled nonlinear oscillators. In the field of photonics, synchronization has been implemented in various laser and oscillator systems, enabling applications including coherent beam combining and high precision pump-probe measurements. Recent experiments have also shown time-domain synchronization of Kerr frequency combs via coupling of two separate oscillators operating in the dissipative soliton [\textit{i.e.}, anomalous group-velocity dispersion (GVD)] regime. Here, we demonstrate all-optical synchronization of Kerr combs in the non-solitonic, normal-GVD regime in which phase-locked combs with high pump-to-comb conversion efficiencies and relatively flat spectral profiles are generated. Our results reveal the universality of Kerr comb synchronization and extend its scope beyond the soliton regime, opening a promising path towards coherently combined normal-GVD Kerr combs with spectrally flat profiles and high comb-line powers in an efficient microresonator platform.
\end{abstract}

\maketitle

Synchronization is a fundamental phenomenon that can occur with coupled nonlinear oscillators and allows for the operating frequencies of different oscillators to become identical \cite{Strogatz2001}. It has been observed throughout nature and across various disciplines of science, where many of its applications have become indispensable. For example, the irregular respiratory rhythm of a patient could be restored through entrainment by mechanical ventilation \cite{Graves1986}, and superconducting Josephson junction arrays can be frequency locked to be utilized as tunable local oscillators at millimeter wavelengths \cite{Wiesenfeld1998}. Within the context of systems exhibiting solitonic behavior, synchronized dissipative solitons in traveling-wave field-effect transistors can be used as an efficient platform for generating phase-controlled (\textit{e.g.}, multiphase) pulse trains \cite{Narahara2018}. In the field of photonics, synchronization has been a key enabler for numerous applications. For example, phase-locked laser arrays allow for coherent beam combining and the enhancement of optical power \cite{Nixon2012}. Synchronized timing between two pulsed lasers \cite{Ma2001} enables pump-probe measurements to operate at the shot-noise limit \cite{Potma2002}, and independent femtosecond lasers can coherently synthesize optical pulses with durations shorter than those achievable from a single laser \cite{Shelton2001}. Additionally, passive, all-optical synchronization schemes of soliton mode-locked lasers reduce complexity and cost \cite{Zhu2005,Tsai2013,Wu2014} and potentially allow for long-term stable operation without the need for temperature stabilization or vibration isolation \cite{Huang2018}.

Recent demonstrations have further extended the realm of synchronization to that of soliton Kerr frequency combs, where the locking of repetition rates or equivalently the comb-line spacings is desirable for many applications. In these experiments, the Kerr combs were operating in the anomalous group-velocity dispersion (GVD) regime in which dissipative cavity solitons are excited as a result of a dual balance between dispersion and nonlinearity, and between loss and parametric gain \cite{Jang2018,Jang2019}. In this regime, a clear analogy between the Kuramoto model of synchronization \cite{Kuramoto1975,Kuramoto1984} and the Lugiato-Lefever equation describing soliton comb dynamics \cite{Lugiato1987} can be identified by approximating the soliton with an analytic expression \cite{Jang2018,Jang2019}. Although such Kerr combs operating in the single-soliton regime can be accessed through simple excitation via pump detuning mechanisms, they exhibit low pump-to-comb conversion efficiencies ($\sim$$1\%$) and follow squared hyperbolic-secant spectral profiles \cite{Bao2014} which may be less favorable for some real-world applications due to the exponential roll-off.

Alternatively, another class of Kerr combs that operate in the normal-GVD regime have been shown to readily offer high pump-to-comb conversion efficiencies ($>$$30\%$) and to exhibit slower spectral power falloffs over the region of interest \cite{Xue2017,Kim2019}. All-optical synchronization of such Kerr combs could allow for coherent beam combining \cite{Kozlov1999} and open a pathway to significantly increase comb-line powers in an efficient microresonator platform. Applications of microresonator-based frequency combs such as data communications \cite{Marin-Palomo2017, Fulop2018} and spectroscopy \cite{Suh2016,Yu2017} could benefit immensely from these high comb-line powers that are otherwise inaccessible without the use of additional amplifiers. For instance, wavelength-division multiplexing (WDM) techniques have employed Kerr combs as a source to replace a laser array and its limited number of channels \cite{Marin-Palomo2017, Fulop2018}. Nevertheless, these techniques required the use of additional bulk amplifiers that would not be needed through the synchronization of normal-GVD Kerr combs. Moreover, substantial advancement in integrating microresonator-based frequency combs has been made by utilizing low-power, electrically-pumped sources \cite{Stern2018,Pavlov2018,Raja2019,Jin2021}. Utilizing such electrically-pumped sources and coherent combining of multiple normal-GVD Kerr combs, higher comb-line powers could be synthesized directly on-chip, leading to a highly-efficient method to produce a large number of high-power channels.

Kerr frequency combs in the normal-GVD regime form through the interlocking of switching waves that connect the two steady-state solutions of the bistable cavity response \cite{Lobanov2015a,Xue2015a,Parra-Rivas2016}. Unlike solitons in the anomalous-GVD regime, normal-GVD combs require an initial perturbation to enable the excitation of switching waves and the comb generation process. The pair of switching waves lock to one another and lead to a phase-locked, low-noise comb state exhibiting intricate temporal and spectral features with high conversion efficiencies and slow spectral power falloffs in the region of interest \cite{Xue2017,Kim2019}. Normal-GVD combs can be generated through pump modulation at the microresonator free spectral range (FSR) \cite{Lobanov2015,Xu2021} or through local perturbations in the dispersion to enable modulation instability, a source of switching-wave excitation \cite{Liang2014,Jang2016}. A common method to create these perturbations is through the coupling between different mode families of a microresonator \cite{Liu2014,Xue2015a,Jang2016,Xue2017,Fulop2018,Kim2019}. The spectral overlap of modes in different mode families can lead to a mode splitting at their degeneracy point \cite{Carmon2008,Ramelow2014} and effectively create a local region of anomalous GVD to allow for modulation instability. It is not known if the theoretical model for coupled normal-GVD combs can be reduced to the Kuramoto model, and thus the possibility that synchronization could occur in such a system has been uncertain.

Here, we show that non-solitonic synchronization of two high-efficiency Kerr combs operating in the normal-GVD regime can be realized. We employ two coupled-ring devices to generate two separate high-efficiency normal-GVD combs and investigate their synchronization properties. Through careful modeling of our system, we accurately describe the influence of coupling between the two devices and theoretically reproduce key features observed in experiment, including a synchronization region known as an Arnold tongue \cite{Jang2019}. Additionally, we demonstrate that two such synchronized combs can be coherently combined to produce even higher comb-line powers for applications such as data communications.

\section{Results}

\subsection{Synchronization scheme and theory}

\begin{figure}[!tb]
  \centering
  \includegraphics[]{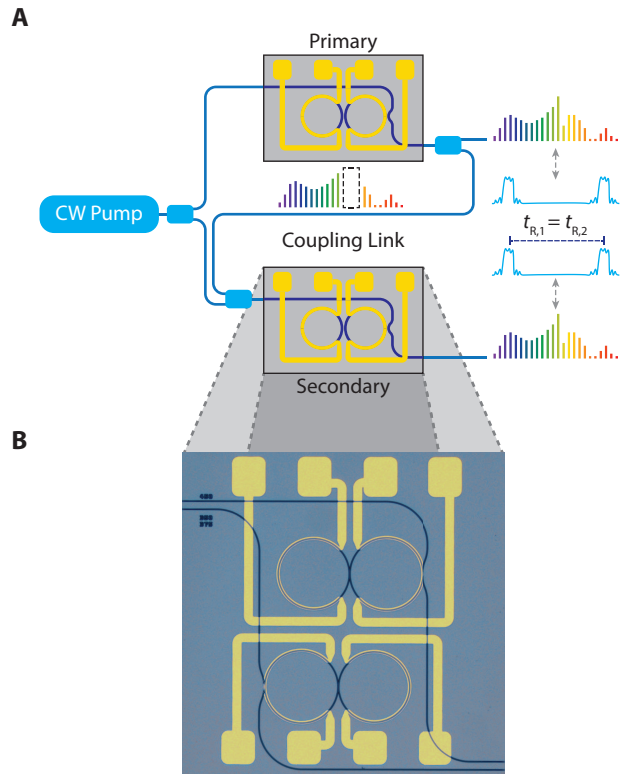}
  \caption{\textbf{Experimental setup.} (\textbf{A})~Schematic of experiment consisting of 2 coupled-ring devices. A small fraction of the output of the primary device is filtered and combined with the pump into the input of the secondary device. The two Kerr combs have roundtrip times, $t_{\mathrm{R},1}$ and $t_{\mathrm{R},2}$, that become identical after synchronization. (\textbf{B})~Microscope image of a single electron-beam lithography field (1$\times$1~mm$^2$) containing two independent Si$_3$N$_4$ coupled-ring devices with integrated platinum resistive heaters. CW, continuous-wave.}
  \label{fig:figure_1}
\end{figure}

A schematic for normal-GVD Kerr comb synchronization is shown in Fig. \hyperref[fig:figure_1]{1A}. For comb generation, we utilize a coupled-ring geometry consisting of a main ring and an auxiliary ring (Fig. \hyperref[fig:figure_1]{1B}) to precisely control an avoided mode crossing produced by the coupling between the two rings \cite{Kim2019}. Two replicas of this device are pumped by a continuous-wave (CW) source and individually tuned to generate separate phase-locked, high-efficiency normal-GVD combs. The light circulates within the main resonator of each device with an associated roundtrip time $t_\mathrm{R}$, which is the inverse of the comb-line spacing and is dependent on the device parameters and environmental fluctuations. Even with state-of-the-art fabrication and typical operating conditions, there is often a small difference in repetition rates ($\Delta f_{\mathrm{rep}}$) between the two combs due to slight fabrication variances and thermal gradients. When phase-locked frequency combs are produced in each device, this small difference in the repetition rates can be measured as a beatnote in the radio frequency (RF) domain when the two optical outputs are combined. We create a coupling link from the primary device to the secondary device (Fig. \hyperref[fig:figure_1]{1A}) in which a small portion of the output from the primary device is combined with the pump of the secondary device. For sufficiently strong coupling between the two devices, the repetition rates synchronize.

\begin{figure*}[!tb]
  \centering
  \includegraphics[]{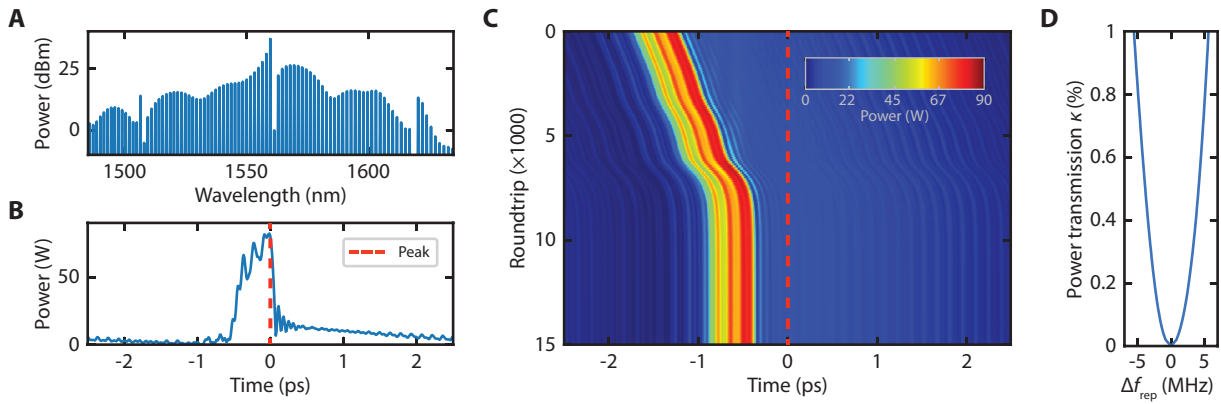}
  \caption{\textbf{Synchronization simulation.} (\textbf{A})~Simulated intracavity optical spectrum of the primary system. (\textbf{B})~Corresponding time-domain trace. The peak is indicated by a red dashed line. (\textbf{C})~Time-domain evolution of secondary structure with respect to the peak of the primary structure (red dashed line). Initial slope corresponds to a 3~MHz relative drift. Due to a fractional power transmitted $\kappa=1\%$, the secondary structure slows down and locks to the position of the primary structure. (\textbf{D})~Synchronization region across various coupling strengths. A slight asymmetry is observed.}
  \label{fig:figure_2}
\end{figure*}

We theoretically study our configuration by implementing a system consisting of a pair of coupled rings. Each coupled-ring model consists of a unitary coupling matrix and a modified nonlinear Schrödinger equation (NLSE) (see Materials and Methods). Figure \hyperref[fig:figure_2]{2A} shows the simulated intracavity spectrum, and Fig. \hyperref[fig:figure_2]{2B} shows the corresponding time-domain trace of our primary normal-GVD comb. We implement the coupling link between the two coupled-ring systems by filtering out the pump line and the three neighboring higher wavelength lines from the primary comb (due to the limited resolution of the 4-$f$ shaper used in the experiment) before feeding a fraction of the filtered output into the input of the secondary coupled-ring system. We simulate a repetition-rate difference corresponding to 3~MHz (see Materials and Methods) and introduce the coupling link with a small fraction ($\kappa$) of the transmitted power of the primary comb equal to $1\%$. Figure \hyperref[fig:figure_2]{2C} plots a density map showing the evolution, from top to bottom, of the secondary time-domain trace with respect to the peak of the primary structure (red dashed line). As the secondary pulse approaches the location of the primary pulse, we observe what appears to be an attracting potential that locks the secondary structure's position to that of the primary structure. Unlike the soliton case, the normal-GVD comb does not have a known analytic expression, and we cannot reduce our model to a dynamical equation equivalent to the Kuramoto model. Nevertheless, we still observe the synchronization of the repetition rates of the two nonlinear oscillators. If the coupling between the two coupled-ring systems is removed, the secondary structure continues to drift at the original injected repetition rate without observing the effects of the primary structure. We characterize the synchronization boundaries for various coupling strengths (\textit{i.e.}, the Arnold tongue) for our system (Fig. \hyperref[fig:figure_2]{2D}). As observed with solitonic synchronization, the synchronization region grows as the fractional coupled power increases. We also observe a slight asymmetry in the region which we attribute to the asymmetry of the coupled spectrum (see Fig. \hyperref[fig:figure_2]{2A}) \cite{Jang2019}.

\subsection{Experimental demonstration}

\begin{figure*}[!tb]
  \centering
  \includegraphics[]{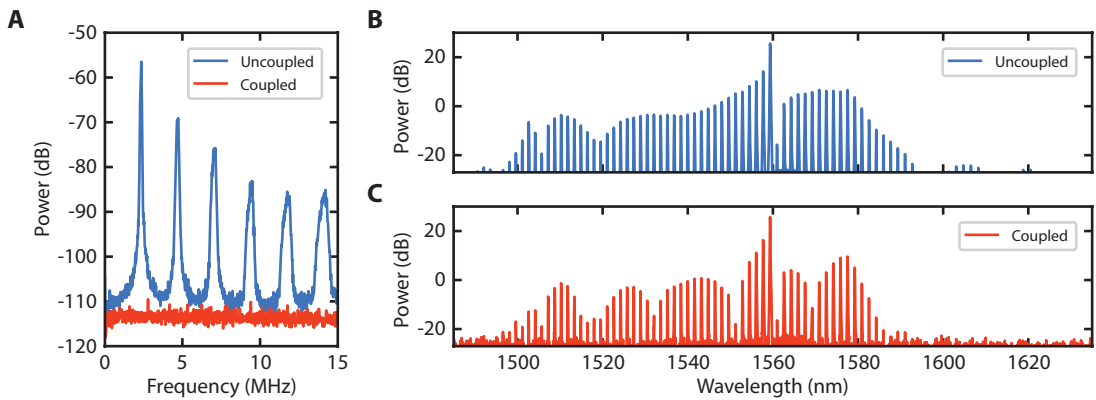}
  \caption{\textbf{Experimental characterization of synchronization.} (\textbf{A})~Measured RF spectra of combined optical outputs. Beatnotes are observed for the uncoupled case. Beatnotes vanish as the coupling is introduced and increased in strength. (\textbf{B})~Measured optical spectrum of combined optical outputs for the uncoupled case. (\textbf{C})~Measured optical spectrum of combined optical outputs for the coupled case. Frequency modulations due to a relative group delay are observed. Differences in comb line powers between the two spectra range from -23 dB to 3 dB.}
  \label{fig:figure_3}
\end{figure*}

Our experimental setup consists of two oxide-clad silicon nitride (Si$_3$N$_4$) coupled-ring devices with integrated platinum resistive heaters (Fig. \hyperref[fig:figure_1]{1B}). The waveguide cross section of our devices is 730$\times$1000 nm$^2$, which allows for a broad region of normal GVD at the pump wavelength for the fundamental TE polarization mode. Each device is pumped with 220 mW of power at 1559.79 nm. The main and auxiliary rings have an FSR of 200 GHz and 206 GHz, respectively. The slight offset in FSR is specifically chosen to exploit the Vernier effect and induce periodic mode interactions spaced by 50 nm since only a single avoided mode crossing is required near the pump wavelength to initiate the comb generation process. In order to precisely control the mode-crossing position, the electrical powers through the heaters for the main and auxiliary rings are individually tuned. Increasing the heater power locally increases the temperature of the respective microresonator and thermally redshifts the associated resonances through the thermo-optic effect. The degeneracy point redshifts (blueshifts) when the smaller (larger) FSR mode family resonances are redshifted. In addition to assisting in the comb-generation process, thermally shifting the main ring resonances allows for fine tuning of the repetition rate or equivalently the frequency spacing of the comb. By exploiting this mechanism, we can tune the relative repetition rate difference between the two combs and explore the region of synchronization for varying coupling strengths. To couple the two devices, we collect a portion of the output of the primary device and filter out the pump using a 4-$f$ shaper. Due to the resolution of the 4-$f$ shaper, the three neighboring higher wavelength lines from the pump are also filtered. The filtered output is then combined with the pump for the secondary device.

Experimental observation of synchronization of two normal-GVD Kerr combs is shown in Fig. \hyperref[fig:figure_3]{3}. Initially, we generate the two combs separately without the coupling link. We offset the primary comb repetition rate by tuning the main ring heater. By combining the outputs, we observe an RF beatnote and its harmonics corresponding to the repetition rate difference between the two combs. For a main ring-heater power of 9.07 mW, we observe a beatnote at 2.36 MHz and the corresponding harmonics (Fig. \hyperref[fig:figure_3]{3A}). We then initiate the unidirectional coupling and observe for a certain threshold coupling strength that the beatnotes vanish, which indicates that the repetition rates of the two combs are now identical and the corresponding comb lines are perfectly overlapped with one another (Fig. \hyperref[fig:figure_3]{3A}). In the optical domain, we observe no modulation in the combined spectrum for the uncoupled case (Fig. \hyperref[fig:figure_3]{3B}) and strong modulation in the combined spectrum for the coupled case (Fig. \hyperref[fig:figure_3]{3C}), which indicates interference of the two sets of comb lines when the repetition rates are synchronized. The differences in comb line powers between the two spectra range from -23 dB to 3 dB. The modulation period is given by the relative group delay when combining the two outputs together. By tuning the path-length difference, we can double the overall power per line or choose to enhance specific sections of the comb. Extending synchronization to multiple microresonator combs would not only allow us to achieve comb-line powers significantly greater than those possible with a single comb, but also provide an approach to shaping the overall spectrum.

\begin{figure}[!tb]
  \centering
  \includegraphics[]{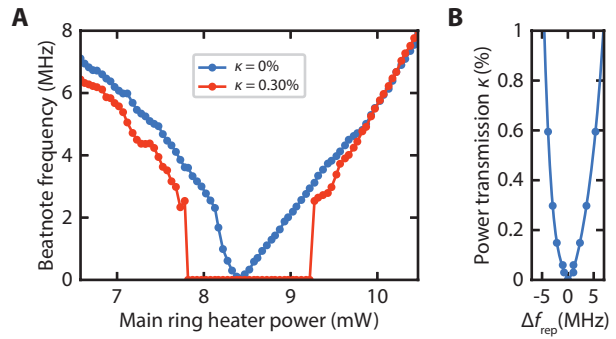}
  \caption{\textbf{Measurement of synchronization region.} (\textbf{A})~Measured beatnotes as main ring heater power is tuned for uncoupled and coupled cases. For $\kappa=0.3\%$, synchronization is observed from 7.8 mW to 9.2 mW of heater power, which corresponds to a synchronization range of -2.9 MHz to 3.6 MHz. (\textbf{B})~Synchronization region across various coupling strengths. Similar to simulation, asymmetry is observed.}
  \label{fig:figure_4}
\end{figure}

We further characterize the synchronization regime by measuring the maximal allowable repetition rate difference for different $\kappa$ values. Figure \hyperref[fig:figure_4]{4A} shows the measured primary beatnotes as we tune the main ring heater power. As a reference, the change in repetition rate with no coupling is shown in red. With a coupling strength of $\kappa=0.3\%$, shown in blue, we observe synchronization from 7.8 mW to 9.2 mW of heater power. Calibrating to the uncoupled data, this range corresponds to a total synchronization range of -2.9 MHz to 3.6 MHz. We then measure this synchronization range for various coupling strengths and obtain a curve known as an Arnold tongue through a fit of the data (Fig. \hyperref[fig:figure_4]{4B}). The measurements are in excellent agreement with our simulations and show the synchronization region increasing with coupling strength. Additionally, the lack of symmetry about $\Delta f_{\mathrm{rep}}=0$ is due to the asymmetric spectrum from the initial source and filtering system \cite{Jang2019}.

\section{Discussion}

Our results reveal the universality of synchronization beyond the dissipative soliton regime by exploring the dynamics of two coupled normal-GVD Kerr combs. Although normal-GVD Kerr combs do not have solitonic profiles, they exhibit synchronization behavior. Moreover, despite the initial perturbation requirements of normal-GVD Kerr combs, key beneficial features such as high pump-to-comb conversion efficiencies make synchronization of such integrated combs desirable. For example, it could allow for coherent beam combining and a dramatic increase in comb-line powers while maintaining high conversion efficiencies. As a passive locking mechanism with low-power requirements, all-optical synchronization can easily be extended to a fully-integrated platform where multiple oscillators can be combined. Such integrated platforms could enhance current WDM schemes where power requirements are not sufficient from single comb generation \cite{Marin-Palomo2017, Fulop2018}. Access to a large number of high-power channels could greatly increase the bandwidth of current data communications protocols. We envision our system would condense complex oscillator networks into a single integrated device.

\section{Materials and Methods}

\subsection{Modeling}

A single coupled-ring model is implemented by mapping the electric field at the beginning of the $(m+1)$th roundtrip to the electric field at the end of the $m$th roundtrip and by using a modified NLSE for propagation. The mapping is given by,
\begin{align}
  \begin{pmatrix}
    E_{\mathrm{out}} \\
    E_1^{(m+1)} \\
    E_2^{(m+1)}
  \end{pmatrix} =
  \mathbf{M}
  \begin{pmatrix}
    E_{\mathrm{in}} \\
    E_1^{(m)} \\
    E_2^{(m)}
  \end{pmatrix}.
\end{align}
Here, $E_{\mathrm{in}}$ is the CW driving field with power $P_\mathrm{in}=|E_{\mathrm{in}}|^2$, $E_1$ is the intracavity field in the main ring, $E_2$ is the intracavity field in the auxiliary ring, $E_{\mathrm{out}}$ is the output field, and $\mathbf{M}=\mathbf{\Theta} \circ \mathbf{A} \circ \mathbf{\Phi}$ is the mapping matrix with $\mathbf{\Theta}$ the unitary coupling matrix, $\mathbf{A}$ the linear attenuation matrix, $\mathbf{\Phi}$ the linear phase accumulation matrix, and $\circ$ the Hadamard product. The unitary coupling matrix $\mathbf{\Theta}$ is given by,
\begin{align}
  \mathbf{\Theta} =
  \begin{pmatrix}
    \sqrt{1 - \theta_1} & i\sqrt{\theta_1(1 - \theta_2)} & -\sqrt{\theta_1\theta_2} \\
    i\sqrt{\theta_1} & \sqrt{(1 - \theta_1)(1 - \theta_2)} & i\sqrt{\theta_2(1 - \theta_1)} \\
    0 & i\sqrt{\theta_2} & \sqrt{1 - \theta_2}
  \end{pmatrix},
\end{align}
where $\theta_1$ is power coupling coefficient between the bus and the main ring and $\theta_2$ is the power coupling coefficient between the main ring and the auxiliary ring. The linear attenuation matrix $\mathbf{A}$ is given by,
\begin{align}
  \mathbf{A} =
  \begin{pmatrix}
    1 & e^{-\frac{\alpha_1 L_1}{2}} & e^{-(\frac{\alpha_1 L_1}{4} + \frac{\alpha_2 L_2}{2})} \\
    1 & e^{-\frac{\alpha_1 L_1}{2}} & e^{-(\frac{\alpha_1 L_1}{4} + \frac{\alpha_2 L_2}{2})} \\
    1 & e^{-\frac{\alpha_1 L_1}{4}} & e^{-\frac{\alpha_2 L_2}{2}}
  \end{pmatrix},
\end{align}
where $\alpha_1$ ($\alpha_2$) and $L_1$ ($L_2$) are the linear attenuation coefficient and cavity length of the main (auxiliary) ring. The linear phase accumulation matrix $\mathbf{\Phi}$ is given by,
\begin{align}
  \mathbf{\Phi} =
  \begin{pmatrix}
    1 & e^{i\phi_1} & e^{i(\frac{\phi_1}{2} + \phi_2)}\\
    1 & e^{i\phi_1} & e^{i(\frac{\phi_1}{2} + \phi_2)} \\
    1 & e^{i\frac{\phi_1}{2}} & e^{i\phi_2}
  \end{pmatrix},
\end{align}
where $\phi_1$ ($\phi_2$) is the roundtrip phase shift of the main (auxiliary) ring.

The nonlinear propagation equations are given by,
\begin{align}
  \frac{\partial E_1}{\partial z} &= \left[i\sum_{k\ge 2}\frac{\beta_k}{k!}\left(i\frac{\partial}{\partial\tau}\right)^k + i\gamma|E_1|^2\right]E_1, \\
  \frac{\partial E_2}{\partial z} &= \left[i\sum_{k\ge 2}\frac{\beta_k}{k!}\left(i\frac{\partial}{\partial\tau}\right)^k - \delta\frac{\partial}{\partial\tau} + i\gamma|E_2|^2\right]E_2, \label{propagation2}
\end{align}
where $z$ is the propagation distance, $\tau$ is the time, $\beta_k$ is the $k$th order dispersion coefficient associated with the Taylor series expansion of the propagation constant $\beta$ about the pump frequency $\omega_0$, and $\gamma$ is the effective nonlinear coefficient. The FSR difference between the main and auxiliary rings is compensated by a linear phase slope across the auxiliary cavity modes $\delta = \frac{1}{L_1}\left(\frac{1}{\mathrm{FSR}_2}-\frac{1}{\mathrm{FSR}_1} \right)$ where FSR$_1$ (FSR$_2$) is the FSR of the main (auxiliary) ring.

Using rounded experimental values and the COMSOL Multiphysics mode-solver package, the following parameters are used: $\alpha_1=8.643$ m$^{-1}$, $\alpha_2=4.321$ m$^{-1}$, $\theta_1=0.01074$, $\theta_2=0.006816$, $L_1=699$~$\mu$m, $L_2=678$~$\mu$m, $P_\mathrm{in}=220$ mW, FSR$_1=200$ GHz, FSR$_2=206$~GHz, $\beta_2=40$~fs$^2$/mm, $\gamma=1.229$~W$^{-1}$m$^{-1}$, $\phi_1=-0.05340$, and $\phi_2=-0.1891$.

Synchronization is investigated by implementing a coupling link between two of the coupled-ring models. The coupling link filters out the pump line and the 3 neighboring higher wavelength lines of the output of the primary model before feeding a fraction of it into the input of the secondary model. The repetition-rate difference $\Delta f_{\mathrm{rep}}$ is simulated by offsetting the group velocity in the secondary coupled-ring model. The group velocity offset parameter $\Delta\beta_1$ is expressed as,
\begin{align}
  \Delta\beta_1 = \frac{t_{\mathrm{R},2} - t_{\mathrm{R},1}}{L_1} \approx \frac{1}{L_1}\frac{\Delta f_{\mathrm{rep}}}{\mathrm{FSR}_1^2}.
\end{align}

\subsection{Experimental setup}

Two independent Si$_3$N$_4$ chips each with a coupled-ring device is pumped with 220 mW of power in the bus waveguide. The pump sources for both coupled-ring devices are derived from a single CW laser at 1559.79 nm with a narrow linewidth of $<$1 kHz. The laser output is amplified by an erbium-doped fiber amplifier and split into 2 light fields with a fused fiber coupler. One field is coupled to the bus waveguide of the primary device using a lensed fiber. Using a non-polarizing beamsplitter, the other field is combined with a filtered and attenuated portion of the primary output. To ensure that the polarization state of the filtered primary output is collinear with that of the secondary pump, a quarter-wave plate and a half-wave plate is utilized. The combined light is coupled to the secondary device using a microscope objective. The unfiltered portion of the primary output and an attenuated secondary output are combined with a non-polarizing beamsplitter before being collected into an optical fiber using a fiber collimator package. The fiber output is then split with a 90/10 fused fiber coupler. Light from the higher power arm is detected by a 10-GHz photodiode, and the RF signal is observed on an RF spectrum analyzer. Light from the lower power arm is used to observe the combined spectrum on an optical spectrum analyzer.
\newline
\newline
\noindent\textbf{Acknowledgments:} This work was performed in part at the Cornell NanoScale Facility, a member of the National Nanotechnology Infrastructure Network, which is supported by the NSF.
\newline
\newline
\noindent\textbf{Funding:} Research was supported by the Advanced Research Projects Agency of the U.S. Department of Energy (grant no. DE-AR0000843), the Air Force Office of Scientific Research (grant no. FA9550-15-1-0303), and
the Defense Advanced Research Projects Agency of the U.S. Department of Defense (grant no. HR0011-19-2-0014).
\newline
\newline
\noindent\textbf{Author contributions:} B.Y.K. performed the experiment, carried out the theoretical analysis and numerical simulation, and wrote the manuscript with inputs from all the authors. B.Y.K. and J.K.J. developed the model for simulations. B.Y.K., J.K.J., Y.O., and A.L.G. contributed to the interpretation of data. X.J. fabricated the devices under the supervision of M.L. A.L.G. supervised the overall project.
\newline
\newline
\noindent\textbf{Competing interests:} The authors declare that they have no competing interests.
\newline
\newline
\noindent\textbf{Data and materials availability:} All data needed to evaluate the conclusions in the paper are present in the paper. Additional data related to this paper may be requested from the authors.

\bibliography{paper}

\end{document}